\documentclass[a4paper,UKenglish,notab,nolineno]{socg-lipics-v2021}
\usepackage{tikz}
\usepackage{etoolbox}

\newcommand{\R}{\mathbb R}

\DeclareMathOperator{\VC}{VCdim}
\DeclareMathOperator{\diam}{diam}

\title{Fixed-Radius Neighbourhoods in Three-Dimensional Unit-Cube Graphs}
\titlerunning{Fixed-Radius Neighbourhoods in Unit-Cube Graphs}
\author{Guangjian Zhang}{\href{mailto:zgj1226029469@outlook.com}{\nolinkurl{zgj1226029469@outlook.com}}}{}{}{}
\authorrunning{Guangjian Zhang}
\Copyright{}
\hideLIPIcs
\EventShortTitle{}
\DOIPrefix{}

\makeatletter
\patchcmd{\endabstract}{%
  \subjclassHeading
  \ifx\@ccsdescString\@empty
    \textcolor{red}{Author: Please fill in 1 or more \string\ccsdesc\space macro}%
  \else
    \@ccsdescString
  \fi
  \vskip\topmattervskip\baselineskip
  \noindent
}{}{}{\PackageError{preprint}{Could not remove subject classification}{The template definition of endabstract has changed.}}
\let\@oddfoot\@empty
\let\@evenfoot\@empty
\renewcommand{\copyrightline}{}
\makeatother
\AtBeginDocument{%
  \hypersetup{
    pdfauthor={Guangjian Zhang},
    pdfsubject={},
    pdfkeywords={Unit-cube graphs, VC-dimension, graph diameter, geometric range searching},
    pdfinfo={ContactEmail={zgj1226029469@outlook.com}}
  }%
}

\begin{document}
\maketitle

\begin{abstract}
We prove Conjecture~4 of Chan, Chang, Gao, Kisfaludi-Bak, Le, and Zheng (SoCG 2026): for every fixed integer $r\ge0$, the radius-$r$ neighbourhoods of a unit-cube graph in $\R^3$ have bounded VC-dimension.
Our bound is $2^{2^{O(r)}}$, independent of the number of vertices.
The main ingredient is a VC-dimension bound for reachability through arbitrary fixed-length sequences of three-dimensional generalized dominance relations.
Its proof uses groups ordered in two coordinates, with neighbouring groups acting as barriers that preserve this invariant when one layer is removed.
Combining these bounds with the existing interval-representation framework, we obtain a Las Vegas algorithm that, for every fixed integer $\Delta\ge1$, decides whether a graph given by $n$ cube centres has diameter at most $\Delta$ in expected $O_\Delta(n^{2-\varepsilon_\Delta})$ time, for an explicit constant $\varepsilon_\Delta>0$.
Finally, for every integer $r\ge7$, we construct unit-cube graphs with rational centres whose radius-$r$ neighbourhoods have VC-dimension $\Omega(r^3)$.
\end{abstract}

\section{Introduction}

The diameter of a graph is the largest distance between two of its vertices.
For geometric intersection graphs, the input is a collection of objects, and two vertices are adjacent when their objects intersect.
Such a graph may have quadratically many edges.
A truly subquadratic algorithm must therefore avoid constructing all edges.

Recent work has established a close connection between diameter computation and the VC-dimension of graph neighbourhoods.
Chang, Gao, and Le~\cite{chang2024diameter} proved a bound of four for the distance VC-dimension of pseudo-disk graphs and used it to obtain an additive approximation for the diameter of unit-disk graphs.
Chan, Chang, Gao, Kisfaludi-Bak, Le, and Zheng~\cite{chan2025vc} subsequently gave truly subquadratic algorithms for exact diameter in unit-disk graphs and intersection graphs of axis-aligned squares.
Their framework combines bounds on VC-dimension with implicit representations of reachable sets.

In three dimensions, the known hardness results leave open the fixed-threshold regime.
Bringmann, Kisfaludi-Bak, K\"unnemann, Nusser, and Parsaeian~\cite{bringmann2022diameter} proved conditional quadratic lower bounds for diameter in unit-cube graphs using a construction whose diameter grows with the input size.
This leaves room for a truly subquadratic algorithm for each fixed threshold.
The authors in~\cite{chan2026landscape} obtained such algorithms for thresholds two and three and proved bounded VC-dimension for radius-two and radius-three neighbourhoods.
Conjecture~4 of their paper asks whether bounded VC-dimension persists for every fixed radius.
We answer this question by giving a reachability bound that applies to an arbitrary fixed number of layers.

\paragraph*{Our results.}
We prove this conjecture.
Throughout the paper, cubes are closed, axis-aligned, and have side length one in $\R^3$.
For a finite set of centres $P$, let $G(P)$ have vertex set $P$ and an edge between distinct $p,q$ precisely when $\|p-q\|_\infty\le1$.
Distances are unweighted graph distances, with distance infinity between different components.
Write
\[
 N^r[v]=\{u:\operatorname{dist}_{G(P)}(u,v)\le r\},
 \qquad
 f^*(r)=\sup_{P\subset\R^3\text{ finite}}
 \VC\bigl(P,\{N^r[v]:v\in P\}\bigr).
\]
A family $\mathcal F$ \emph{shatters} a set $X$ if its traces $\{F\cap X:F\in\mathcal F\}$ comprise every subset of $X$.
Its VC-dimension is the largest size of a shattered set.
Throughout, the radius is fixed when defining the neighbourhood set system.

\begin{theorem}[Resolution of Conjecture 4 in~\cite{chan2026landscape}]\label{thm:vc}
For every fixed integer $r\ge0$, the radius-$r$ neighbourhoods of three-dimensional unit-cube graphs have bounded VC-dimension.
More precisely, $f^*(r)\le2^{2^{O(r)}}$, independently of the number of vertices.
\end{theorem}

The bound yields an algorithm for every fixed diameter threshold.

\begin{theorem}\label{thm:diameter}
For every fixed integer $\Delta\ge1$, there is a Las Vegas algorithm that, given $n$ cube centres, decides whether $\diam(G(P))\le\Delta$ in expected time
\[
 O_\Delta\bigl(n^{2-\varepsilon_\Delta}\bigr)
\]
on the real RAM, for an explicit constant $\varepsilon_\Delta>0$.
\end{theorem}

After handling $n\le1$ directly, applying this decision algorithm to the thresholds $1,\ldots,\Delta$ computes the exact diameter of graphs whose diameter is at most a fixed constant $\Delta$.
Our results extend the bounded-VC-dimension and truly subquadratic decision results of~\cite{chan2026landscape} from radii and thresholds two and three to every fixed radius and threshold.
For thresholds two and three, the specialized algorithms in~\cite{chan2026landscape} are faster.
Our final result quantifies the dependence of VC-dimension on the radius.

\begin{theorem}\label{thm:lower}
For every integer $r\ge7$, there exists a unit-cube graph with rational cube centres containing a set of
\[
 k=8^{\lfloor\log_2((r+5)/6)\rfloor}
\]
vertices shattered by its radius-$r$ neighbourhoods.
Consequently, $f^*(r)\ge k>(r+5)^3/1728$.
\end{theorem}

\paragraph*{Explicit bounds.}
For $r\ge1$, define
\[
 h_r=6^{8(2^r-1)},\qquad
 e_r=27^r h_r+\left\lceil\log_2((2r+1)^3)\right\rceil+2.
\]
The proof of Theorem~\ref{thm:vc} gives $f^*(r)\le2e_r^2-1$.
Also, $f^*(0)=1$.
For Theorem~\ref{thm:diameter}, we may take
\[
 d_\Delta=\left(\sum_{j=0}^{\Delta}27^j\right)h_\Delta,
 \qquad \varepsilon_\Delta=\frac{1}{4(d_\Delta+1)}.
\]
For example, $f^*(4)\le2^{339}-1$, and threshold four can be decided in expected time $O(n^{2-2^{-332}})$.
Appendix~\ref{ap:numeric} gives the arithmetic for these explicit estimates.

\paragraph*{Proof ideas.}
Our main technical contribution is a VC-dimension bound for reachability through arbitrary fixed-length sequences of generalized dominance relations in three dimensions.
The grid reduction of~\cite{chan2026landscape} turns cube intersection between two fixed unit cells into such a relation.
A prescribed cell path then gives a layered system, with one relation for each step.
We prove a stronger statement for groups of starting points that are strictly ordered in two coordinates.
This formulation lets us remove the first layer and treat the possible successors as new groups.

In each group, we choose a point minimizing the remaining coordinate.
After passing to a monotone subsequence of representatives, we retain the even groups.
The odd groups act as barriers: if none of them can reach an endpoint, a first step on a path to that endpoint must lie in two specified coordinate intervals.
The relevant successors are covered by three fixed families, one for each coordinate pair.
Each family has two ordered coordinates, so induction and trace counting apply.
Section~\ref{sec:upper} gives the full argument and the reduction to cube neighbourhoods.

Applying the interval-representation approach of~\cite{chan2025vc} with these reachability bounds gives the fixed-threshold diameter algorithm.
For each starting cell, one order of its vertices serves all cell paths and all their prefixes.
The reversed reachability bound controls membership patterns and hence the total interval complexity in this order.
Coloured orthogonal range queries propagate the intervals exactly.
Section~\ref{sec:algorithm} gives the implementation and the explicit exponent, with the data structures proved in Appendix~\ref{ap:algorithm}.

For the lower bound, we use the label-encoding mechanism of~\cite{bringmann2022diameter} on a three-dimensional branching tree with many equal-depth leaves.
At scale $h$, this tree has $8^h$ leaves and depth $O(2^h)$.
The labels produce one witness for each subset of the leaves, and the common depth makes all witnesses use the same radius.
Section~\ref{sec:lower} explains the geometric realization and the exact distances.
Appendix~\ref{ap:lower} gives its complete verification.

\section{Reachability through dominance relations}\label{sec:upper}

We first study paths whose successive edges are defined by coordinate comparisons.
These are the generalized dominance relations used in~\cite{chan2026landscape}.
The main step is an induction for groups of starting points that are ordered in two coordinates.

\subsection{Layered systems and ordered groups}

A \emph{layered system of length $t$} consists of finite sets $L_0,\ldots,L_t\subset\R^3$ and relations $\prec_j$ between $L_{j-1}$ and $L_j$, for $1\le j\le t$.
Each relation is the conjunction of three coordinate conditions.
In each coordinate, the condition is one of $<$, $>$, $\le$, $\ge$, or the condition that is always true, denoted by $\top$.
The choices are fixed for each relation.
Points in different layers are regarded as distinct copies, even if their coordinates agree.
We write $p\leadsto s$ if there are points $p_j\in L_j$ with $p_0=p$, $p_t=s$, and $p_{j-1}\prec_j p_j$ for every $j$.
For $t=0$, this means $p=s$.

Let $Q_1,\ldots,Q_k\subseteq L_0$ be fixed groups of points.
They have \emph{two ordered coordinate bands} if there are distinct coordinates $a,b$ with the following property.
For each $d\in\{a,b\}$, either
\[
 q_d<q'_d\quad\text{for all $i<j$, $q\in Q_i$, and $q'\in Q_j$},
\]
or all these inequalities are reversed.
The directions in coordinates $a$ and $b$ may differ.
Empty groups are allowed.
Each endpoint $s\in L_t$ defines a range on the group labels,
\begin{equation}\label{up:group-range}
 F_s=\{i\in[k]:\text{some $q\in Q_i$ satisfies $q\leadsto s$}\}.
\end{equation}

We use three elementary facts.
The first permits us to work with strict comparisons.
The other two provide the counting step of the induction: the Sauer--Shelah--Perles bound~\cite{sauer1972density,shelah1972stability} and the Erd\H{o}s--Szekeres monotone subsequence lemma~\cite{erdos1935geometry}.
Coordinate normalization and the classical trace bound are proved in Appendices~\ref{ap:normalization} and~\ref{ap:count}.
We give the remaining counting arguments here.

\begin{lemma}[Coordinate normalization]\label{up:normalization}
A finite layered system can be represented so that the point values in each coordinate are distinct and every coordinate condition is strict or $\top$, while preserving every adjacent-layer relation.
Any prescribed strict coordinate orders within a layer can also be preserved.
If $t\ge1$, we can further make the first relation strict dominance,
$q\prec_1 w$ if and only if $q_d<w_d$ for all three coordinates $d$.
This last operation preserves ordered bands, possibly reversing their directions.
\end{lemma}

The lemma has two stages.
The perturbation preserves prescribed strict coordinate orders.
The subsequent normalization of the first relation uses translations and reflections.
It preserves the ordered-band property, although a reflection may reverse its direction.
Both stages preserve every adjacent-layer relation, and hence every layered path.

\begin{lemma}[Trace counting]\label{up:trace-counting}
Let $g\ge0$ be an integer.
A set family of VC-dimension at most $g$ has at most
$\sum_{j=0}^{g}\binom mj\le(m+1)^g$ traces on any $m$ elements.
If $g\ge1$, the unions obtained by choosing one set from each of three fixed families of VC-dimension at most $g$ do not shatter $18g$ elements.
\end{lemma}

\begin{proof}[Proof of the three-family assertion]
Fix a sample of $m=18g$ elements.
Since $18^{-j}\ge18^{-g}$ for $j\le g$, the trace bound and the binomial theorem give at most
\[
 \sum_{j=0}^g\binom{18g}{j}
 \le18^g\left(1+\frac1{18}\right)^{18g}
 <54^g
\]
traces for each family.
For the strict inequality, use
\[
 \left(1+\frac1{18}\right)^{18}
 \le\sum_{j=0}^{18}\frac1{j!}<3.
\]
Indeed, $j!\ge2^{j-1}$ for $j\ge2$, with strict inequality for $j\ge3$, and $\sum_{j=2}^{\infty}2^{-(j-1)}=1$.
A union of three sets is determined by the ordered triple of their traces.
The three families therefore produce fewer than $54^{3g}$ unions on the sample.
Since $54^3=157464<262144=2^{18}$, this is less than $2^{18g}$.
Hence the unions cannot shatter the sample.
\end{proof}

\begin{lemma}[Monotone subsequences]\label{up:monotone}
For every integer $s\ge1$, a sequence of more than $(s-1)^2$ distinct real numbers has a strictly increasing or strictly decreasing subsequence of length $s$.
\end{lemma}

\begin{proof}
At each position, record the lengths of the longest increasing and decreasing subsequences ending there.
Consider two positions, with the first preceding the second.
If the second value is larger, an increasing subsequence ending at the first position extends to the second, so the increasing length strictly grows.
If the second value is smaller, the decreasing length strictly grows instead.
The pairs at distinct positions are therefore distinct.
If no monotone subsequence has length $s$, both entries belong to $\{1,\ldots,s-1\}$.
There would then be at most $(s-1)^2$ positions, a contradiction.
\end{proof}

\subsection{The induction on two ordered bands}

Taking the first step replaces a starting point by a set of possible successors.
We therefore formulate the induction for groups of starting points.
The invariant is that the groups are strictly ordered in two coordinates.
In three dimensions this leaves one free coordinate, in which we choose a minimizing representative from each group.
The barrier argument covers the relevant successors by three fixed families, one for each coordinate pair.
Each family satisfies the same invariant after the first layer is removed.

Define $g_0=1$ and $g_t=1296g_{t-1}^2$ for $t\ge1$.
The resulting closed forms are
\begin{equation}\label{up:constants}
 g_t=6^{4(2^t-1)}
 \qquad\text{and}\qquad
 h_t=g_t^2=6^{8(2^t-1)}.
\end{equation}

\begin{lemma}\label{up:two-bands}
In a layered system of length $t$, let $Q_1,\ldots,Q_k$ have two ordered coordinate bands.
Then the family of label ranges in~\eqref{up:group-range} has VC-dimension at most $g_t$.
\end{lemma}

\begin{proof}
We proceed by induction on $t$.
When $t=0$, different nonempty groups are disjoint because they are strictly ordered in a coordinate.
An endpoint therefore belongs to at most one group.
Every label range has size at most one, so the VC-dimension is at most $1=g_0$.

Suppose that $t\ge1$ and the assertion holds for $t-1$.
Write $g=g_{t-1}$ and $m=18g$.
Assume for a contradiction that a set of $K$ labels is shattered, where
\begin{equation}\label{up:shattered-size}
 K=(2m)^2+1=1296g^2+1.
\end{equation}
We restrict to those labels and retain their order.
All their groups are nonempty.
Apply Lemma~\ref{up:normalization} so that the first relation is strict dominance and all coordinates are distinct.
The two ordered bands remain ordered.
Let $c$ denote the third coordinate.

\paragraph*{Representatives and barriers.}
For each $i$, choose a representative $q_i^*\in Q_i$ minimizing coordinate $c$.
This representative is an actual point of $Q_i$.
Its minimum property has the following useful direction: if $u_c<(q_i^*)_c$, then $u_c<q_c$ for every $q\in Q_i$.
The values $(q_i^*)_c$ are distinct.
By Lemma~\ref{up:monotone} and~\eqref{up:shattered-size}, they have a monotone subsequence of length $2m+1$.
The representatives in this subsequence are also strictly monotone in each of the other two coordinates, by the ordered-band condition.
Restrict to this subsequence and relabel its groups as $Q_1,\ldots,Q_{2m+1}$.
Each of the three coordinate sequences of representatives is now strictly monotone.

We keep the even groups and use the odd groups as barriers.
The retained label $i\in[m]$ represents the original group $Q_{2i}$.
For every $A\subseteq[m]$, shattering gives an endpoint $s_A$ whose trace on the selected $2m+1$ labels is exactly $\{2i:i\in A\}$.
In particular, no odd group reaches $s_A$.
It is this nonreachability that makes the odd groups barriers: their representatives cannot be predecessors of a point on a path to $s_A$.
For each retained label $i$, let
\begin{equation}\label{up:barriers}
 u_i=q_{2i-1}^*,\qquad v_i=q_{2i+1}^*,\qquad
 J_{i,d}=\bigl(\min\{(u_i)_d,(v_i)_d\},
                    \max\{(u_i)_d,(v_i)_d\}\bigr]
\end{equation}
for every coordinate $d$.
For a fixed coordinate, the intervals $J_{1,d},\ldots,J_{m,d}$ are strictly ordered in one direction.
Consecutive intervals share an endpoint, but one contains that endpoint and the other does not.
Thus any point in one interval is strictly ordered with respect to every point in the next interval.

\paragraph*{Recovering two ordered coordinate bands.}
We next show that a path from a retained group to $s_A$ enters at least two of these bands in its first step.
Let $i\in A$ and consider such a path, with first step $q\prec_1 w$, where $q\in Q_{2i}$ and $w\in L_1$.
Let $I_+$ be the coordinates in which the representatives increase with their labels, and let $I_-$ be the coordinates in which they decrease.
For $d\in I_+$, we have
\begin{equation}\label{up:lower-increasing}
 (u_i)_d<q_d<w_d.
\end{equation}
For an ordered-band coordinate, the first inequality holds for every point of $Q_{2i}$.
For the free coordinate $c$, it follows from
$(u_i)_c<(q_{2i}^*)_c\le q_c$.
Similarly, for $d\in I_-$ we have
\begin{equation}\label{up:lower-decreasing}
 (v_i)_d<q_d<w_d.
\end{equation}
Here the free-coordinate case uses
$(v_i)_c<(q_{2i}^*)_c\le q_c$.

Neither $u_i\prec_1 w$ nor $v_i\prec_1 w$ can hold.
Indeed, either relation would extend along the suffix of our path to give a path from an odd group to $s_A$.
By~\eqref{up:lower-increasing}, the failed comparison for $u_i$ must occur in a coordinate $d\in I_-$.
For that coordinate,
\[
 (v_i)_d<w_d\le(u_i)_d,
 \qquad\text{so}\qquad w_d\in J_{i,d}.
\]
By~\eqref{up:lower-decreasing}, the failed comparison for $v_i$ must occur in a coordinate $e\in I_+$.
It gives
\[
 (u_i)_e<w_e\le(v_i)_e,
 \qquad\text{so}\qquad w_e\in J_{i,e}.
\]
The coordinates $d$ and $e$ are distinct.
If either $I_+$ or $I_-$ were empty, one of the two required failed comparisons would be impossible, so the path could not exist.
We have proved that every such first-step endpoint lies in at least two of its three intervals $J_{i,d}$.
Figure~\ref{fig:barrier-bands} shows the two coordinate constraints.

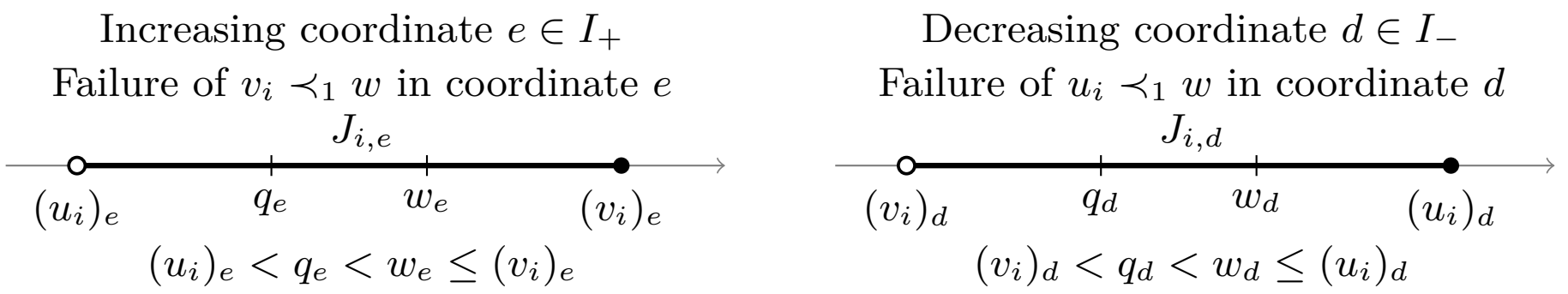
\begin{figure}[tb]
\centering
\begin{tikzpicture}[x=1cm,y=1cm,font=\small]
  \begin{scope}
    \node at (3,1.03) {Increasing coordinate $e\in I_+$};
    \node at (3,.63) {Failure of $v_i\prec_1 w$ in coordinate $e$};
    \draw[->,gray] (.25,0)--(5.8,0);
    \draw[line width=1.2pt] (.8,0)--(5,0);
    \draw[fill=white,line width=.7pt] (.8,0) circle (1.8pt);
    \fill (5,0) circle (1.8pt);
    \draw (2.3,-.08)--(2.3,.08);
    \draw (3.5,-.08)--(3.5,.08);
    \node[below] at (.8,-.06) {$(u_i)_e$};
    \node[below] at (2.3,-.06) {$q_e$};
    \node[below] at (3.5,-.06) {$w_e$};
    \node[below] at (5,-.06) {$(v_i)_e$};
    \node at (3,.25) {$J_{i,e}$};
    \node at (3,-.77) {$(u_i)_e<q_e<w_e\le(v_i)_e$};
  \end{scope}
  \begin{scope}[xshift=6.4cm]
    \node at (3,1.03) {Decreasing coordinate $d\in I_-$};
    \node at (3,.63) {Failure of $u_i\prec_1 w$ in coordinate $d$};
    \draw[->,gray] (.25,0)--(5.8,0);
    \draw[line width=1.2pt] (.8,0)--(5,0);
    \draw[fill=white,line width=.7pt] (.8,0) circle (1.8pt);
    \fill (5,0) circle (1.8pt);
    \draw (2.3,-.08)--(2.3,.08);
    \draw (3.5,-.08)--(3.5,.08);
    \node[below] at (.8,-.06) {$(v_i)_d$};
    \node[below] at (2.3,-.06) {$q_d$};
    \node[below] at (3.5,-.06) {$w_d$};
    \node[below] at (5,-.06) {$(u_i)_d$};
    \node at (3,.25) {$J_{i,d}$};
    \node at (3,-.77) {$(v_i)_d<q_d<w_d\le(u_i)_d$};
  \end{scope}
\end{tikzpicture}
\caption{The two failed comparisons constrain coordinates of the first-step endpoint $w$.
Here $u_i\in Q_{2i-1}$, $q\in Q_{2i}$, and $v_i\in Q_{2i+1}$.
Open and filled endpoints indicate the intervals $(\min,\max]$.
The picture represents coordinate inequalities, not physical obstacles to a path.}
\label{fig:barrier-bands}
\end{figure}

\paragraph*{Fixed successor groups and trace counting.}
Different first-step endpoints may require different coordinate pairs.
We therefore use all three families.
They need not be disjoint.
For each coordinate pair $B\in\{\{x,y\},\{x,z\},\{y,z\}\}$, define
\begin{equation}\label{up:successor-groups}
 Q_i^B=\{w\in L_1:\text{some $q\in Q_{2i}$ satisfies $q\prec_1 w$,
                  and $w_d\in J_{i,d}$ for every $d\in B$}\}.
\end{equation}
These groups depend only on the original groups, their representatives, and the first relation.
They are independent of $A$ and $s_A$, so the induction hypothesis will apply to three fixed set families.
For a fixed $B$, the groups $Q_1^B,\ldots,Q_m^B$ have two ordered bands, one in each coordinate of $B$.
Let $U_B(s)\subseteq[m]$ be their label range after the remaining $t-1$ relations.
We claim that for every $A\subseteq[m]$,
\begin{equation}\label{up:three-union}
 A=U_{\{x,y\}}(s_A)\cup U_{\{x,z\}}(s_A)\cup U_{\{y,z\}}(s_A).
\end{equation}
For the inclusion from right to left, take a path from some $Q_i^B$ to $s_A$.
Its starting point has a predecessor in $Q_{2i}$ by~\eqref{up:successor-groups}.
Prepending that edge shows that $Q_{2i}$ reaches $s_A$, and hence $i\in A$.
For the other inclusion, take a path from $Q_{2i}$ to $s_A$ with $i\in A$.
The preceding barrier argument places its first-step endpoint in $Q_i^B$ for at least one pair $B$.
The suffix then shows that $i\in U_B(s_A)$.

By induction, each of the three fixed families $\{U_B(s):s\in L_t\}$ has VC-dimension at most $g$.
Lemma~\ref{up:trace-counting} says that their unions cannot shatter the $m=18g$ retained labels.
This contradicts~\eqref{up:three-union}.
The VC-dimension is consequently at most $1296g^2$.
Finally, $g_0=1$ and
$1296g_{t-1}^2=6^4\cdot6^{8(2^{t-1}-1)}=6^{4(2^t-1)}=g_t$,
which completes the induction.
\end{proof}

\subsection{Arbitrary starting points}

\begin{theorem}\label{up:layered}
For every layered system of length $t$, the family
$\{\{p\in L_0:p\leadsto s\}:s\in L_t\}$ has VC-dimension at most $h_t$.
It has at most $(k+1)^{h_t}$ traces on any $k$ starting points.
The same bounds hold when all layers and relations are reversed.
In particular, the forward-reachability family
\[
 \bigl\{\{s\in L_t:p\leadsto s\}:p\in L_0\bigr\}
\]
on ground set $L_t$ also has VC-dimension at most $h_t$ and at most $(k+1)^{h_t}$ traces on any $k$ endpoints.
\end{theorem}

\begin{proof}
For $t=0$, every range is a singleton, and $h_0=1$ suffices.
For $t\ge1$, apply the coordinate perturbation in Lemma~\ref{up:normalization}.
Suppose that $g_t^2+1$ starting points are shattered.
Sort them by coordinate $x$.
Their $y$ coordinates are distinct, so Lemma~\ref{up:monotone} gives a subsequence of $g_t+1$ points that is monotone in coordinate $y$.
Make each of these points a singleton group.
The groups have two ordered bands, in coordinates $x$ and $y$, and their labels remain shattered.
This contradicts Lemma~\ref{up:two-bands}.
The trace bound follows from Lemma~\ref{up:trace-counting}, with the empty sample contributing one trace.
Reversing the layer order and reversing each coordinate comparison gives another layered system of length $t$.
Applying the same argument to that system proves the final assertion.
\end{proof}

The two orientations count different patterns.
With endpoints indexing ranges on $L_0$, we count how a fixed sample of starting points reaches an endpoint.
With starting points indexing ranges on $L_t$, we count the membership patterns of starting points in a fixed sample of endpoint ranges.
The latter form will be used to order starting points in Section~\ref{sec:algorithm}.

\subsection{Unit-cube neighbourhoods}\label{up:grid}

We now apply Theorem~\ref{up:layered} to the graph $G(P)$.

\begin{theorem}\label{up:cube-traces}
Let $r\ge1$ be an integer and put
\begin{equation}\label{up:er}
 e_r=27^r h_r+\left\lceil\log_2((2r+1)^3)\right\rceil+2.
\end{equation}
For every finite $P\subset\R^3$ and every $X\subseteq P$ of size $k\ge1$,
\[
 \bigl|\{N^r[v]\cap X:v\in P\}\bigr|\le(k+1)^{e_r}.
\]
\end{theorem}

\begin{proof}
Partition $\R^3$ into the half-open cells $\alpha+[0,1)^3$, for $\alpha\in\mathbb Z^3$.
Write each point as $p=\alpha_p+u_p$, where $u_p\in[0,1)^3$.
If $\|p-q\|_\infty\le1$, then the cell indices differ by at most one in every coordinate.
Fix two such cells and a coordinate $d$.
If $(\alpha_q)_d=(\alpha_p)_d+1$, the intersection condition in coordinate $d$ is $(u_p)_d\ge(u_q)_d$.
If the cell index decreases by one, it is $(u_p)_d\le(u_q)_d$.
If the indices agree, the condition always holds.
Thus intersection between any fixed pair of cells is a generalized dominance relation on the fractional coordinates.
Weak inequalities retain intersections on cube boundaries.

A \emph{cell path} of length $r$ is a sequence of $r+1$ cells whose consecutive indices differ by at most one in each coordinate.
Cells may repeat, and consecutive cells may coincide.
For a fixed cell path, make one layer from the points in each cell and use these relations between consecutive layers.
A repeated point in consecutive layers represents a stationary step.
Any graph path of length at most $r$ can be extended to $r$ steps by staying at its final vertex.
Conversely, deleting stationary steps from a layered path gives a graph walk of length at most $r$.
Thus the union over all cell paths describes precisely the radius-$r$ neighbourhood.

Fix the endpoint cell.
There are at most $27^r$ cell paths ending there, since each preceding cell has at most $27$ possible indices.
For one such path, apply Theorem~\ref{up:layered} to the sample points in its first cell.
It gives at most $(k+1)^{h_r}$ traces.
If that cell contains no sample point, it contributes only the empty trace.
Each endpoint's neighbourhood trace is determined by one trace for each cell path.
The number of traces obtained from this endpoint cell is therefore at most
\begin{equation}\label{up:one-cell-traces}
 (k+1)^{27^r h_r}.
\end{equation}

An endpoint with a nonempty trace is at graph distance at most $r$ from a sample point.
Its cell index differs from that sample point's cell index by at most $r$ in each coordinate.
There are consequently at most $(2r+1)^3k$ endpoint cells with a nonempty trace.
All remaining cells contribute just the empty trace.
With $a=27^rh_r$ and $b=\lceil\log_2((2r+1)^3)\rceil$, we obtain
\[
 \bigl|\{N^r[v]\cap X:v\in P\}\bigr|
 \le1+(2r+1)^3k(k+1)^a
 \le(k+1)^{a+b+2}.
\]
For the last inequality, use $(2r+1)^3\le2^b\le(k+1)^b$, $k\le k+1$, and
$1+(k+1)^{a+b+1}\le(k+1)^{a+b+2}$.
This is the claimed bound.
\end{proof}

\paragraph*{From traces to VC-dimension.}
For an integer $e\ge1$, put $k=2e^2$.
The inequality $2^{e-1}\ge e$ follows by induction: it is an equality for $e=1$, and doubling gives $2^e\ge2e\ge e+1$.
Consequently,
\[
 \log_2(2e^2+1)<\log_2(4e^2)=2+2\log_2 e\le2e.
\]
Multiplying by $e$ gives $e\log_2(k+1)<k$, and hence $(k+1)^e<2^k$.
Theorem~\ref{up:cube-traces} with $e=e_r$ therefore rules out a shattered set of size $2e_r^2$.
This proves $f^*(r)\le2e_r^2-1$ and resolves Conjecture~4 of~\cite{chan2026landscape}.
For radius zero, every neighbourhood is a singleton, so two points cannot be shattered.
A graph with two vertices can shatter one point, giving $f^*(0)=1$.
For $r=4$, using $h_4=6^{120}$ gives the sharper bound $2^{339}-1$.
Appendix~\ref{ap:numeric} verifies this estimate and the corresponding algorithmic exponent.
Lemma~\ref{up:normalization} is used in the structural proof.
The algorithm below works with the original weak comparisons derived from the grid cells, so intersections on cube boundaries are retained.

\section{A subquadratic algorithm for every fixed diameter threshold}
\label{sec:algorithm}

Fix an integer $\Delta\ge 1$.
We apply the interval-representation approach of~\cite{chan2025vc} using the
reachability bounds from Section~\ref{sec:upper}.
Low-crossing orders are a standard tool in geometric range searching~\cite{welzl1988partition}.
The analysis must control intermediate reachable sets as well as final
neighbourhoods.
For each starting cell, we therefore order its vertices once for the labelled
ranges of all cell paths, including their prefixes.
The total interval bound then accounts for the representations used at every
stage of the computation.
Write $\widetilde O$ for a bound that suppresses polylogarithmic factors, and set
\[
 c_\Delta=\sum_{j=0}^{\Delta}27^j,
 \qquad d_\Delta=c_\Delta h_\Delta,
 \qquad \alpha=\frac{1}{d_\Delta+1}.
\]
Constants in this section may depend on $\Delta$.

We first state two tools that follow this approach.
We give self-contained forms for the weak generalized dominance relations
and labelled ranges used here.
Their full proofs and the range-searching structures appear in
Appendix~\ref{ap:algorithm}.
An integer interval is represented by its two endpoints.
The first tool computes each next-layer range from its predecessors without
enumerating all pairs of points satisfying the relation.

\begin{lemma}[Propagating interval representations]
\label{alg:interval}
Let $\lhd$ be a fixed generalized dominance relation in $\R^3$ whose
coordinate conditions belong to $\{\le,\ge,\top\}$.
There are $N$ descriptors $(p,I_p)$, where $I_p\subseteq[1,a]$ is an integer
interval, and $m$ query points.
Let $L$ be the sum of the input interval lengths and $K'$ the total number of
maximal intervals in the sets
$U(q)=\bigcup_{p\lhd q}I_p$.
For any integer $b\ge1$, all these sets can be computed in
\[
 \widetilde O\left((N+m+K')b+\frac{L+ma}{b}+1\right)
\]
time.
\end{lemma}

To see the role of $b$, divide the ordering into blocks of $b$ positions.
For an interval that covers a whole block, store its point in the block's
dominance structure.
A query tests whether at least one such interval contributes to $U(q)$.
Within a partially covered block, give each position its own colour.
A coloured dominance query tests whether the contributing intervals cover
every colour in that block.
Each input interval contributes to at most two partial blocks, using
$O(b)$ copies of its point.
The full blocks account for $O(N+L/b)$ copies in total.
For a query interval that cuts a block, test its positions individually.
Only the two end blocks require these tests.
These structures test whether an interval is covered by $U(q)$.
A separate orthogonal query tests whether it is disjoint from $U(q)$.
Starting from $[1,a]$, split an interval only when neither test succeeds.
The resulting covered intervals yield the exact union.
Appendix~\ref{ap:interval-proof} gives the construction and charges the
splits to boundaries of output intervals.

The second tool constructs one ordering in which the total interval
complexity of a labelled range family is small in expectation.

\begin{lemma}[A common ordering]
\label{alg:ordering}
Let $X$ be a set of at most $n$ elements, and let $\mathcal F$ contain $m$
labelled ranges on $X$.
Suppose any $s$ ranges induce at most $(s+1)^d$ distinct membership vectors
on $X$, and one range can be listed in $\widetilde O(n)$ time.
For any integer $t\ge1$, an ordering of $X$ can be constructed in
$\widetilde O(nt)$ time such that the total number $K$ of maximal intervals
representing the labelled ranges satisfies
\[
 \mathbb E[K]=\widetilde O\left(m\left(1+\frac nt+(t+1)^d\right)\right).
\]
Equal ranges with different labels are counted separately.
\end{lemma}

When $m\ge1$, the construction samples $t$ range labels independently with
replacement.
It groups points with the same membership vector on the sampled ranges and
places each group consecutively.
There are at most $(t+1)^d$ groups.
For any fixed pair of points, let $c$ be the number of original range labels
that separate them.
The probability that the sampled membership vectors agree is
$(1-c/m)^t\le\exp(-tc/m)$.
A pair with $c>6m\log(n+1)/t$ therefore agrees with probability at most $(n+1)^{-6}$.
Taking a union bound over at most $n^2$ pairs, with probability at least
$1-(n+1)^{-4}$ every within-group adjacency crosses at most $6m\log(n+1)/t$ range labels.
There are at most $n$ such adjacencies and at most $(t+1)^d-1$ adjacencies between groups.
Each between-group adjacency crosses at most $m$ range labels.
Since a range has at most one more maximal interval than crossings, this event gives
\[
 K\le m\left(1+\frac{6n\log(n+1)}t+(t+1)^d\right).
\]
For every sample, $K\le mn$.
The complementary event thus contributes at most $mn(n+1)^{-4}\le m$ to the expectation.
This proves the stated expected interval bound.
Appendix~\ref{ap:ordering-proof} gives the full implementation and treats empty families.

\paragraph*{Cell paths and their ranges.}
If $n\le1$, we return yes.
If the spread in any coordinate exceeds $\Delta$, we return no, since one
edge changes each coordinate by at most one.
Otherwise, place the origin of a half-open unit grid at the coordinatewise
minimum of $P$.
At most $(\Delta+1)^3$ cells contain input points.
Cell assignment uses comparisons with $\Delta+1$ integer offsets of the origin.

Fix a nonempty starting cell $B$, and put $X=P\cap B$ and $a=|X|$.
Enumerate all cell paths $\omega$ starting at $B$ with at most $\Delta$ steps.
Each step changes each cell index by $-1$, $0$, or $1$, so there are at most
$c_\Delta$ paths.
For an endpoint $v$ in the last cell of $\omega$, let $F_{\omega,v}$ be the
set of points of $X$ that reach $v$ along this cell path.
Use the exact weak comparisons from Section~\ref{up:grid}.
The length-zero ranges are the singletons of $X$.
Together these labelled ranges form a family $\mathcal F_B$ of size
$m\le c_\Delta n$.
Labels are pairs $(\omega,v)$, so equal sets arising from different paths
remain separate members of the family.
Every prefix is itself one of the enumerated paths and contributes its
endpoint ranges to this same family.

Consider $s$ selected range labels.
Let $s_\omega$ be the number of selected range labels associated with $\omega$.
For a fixed path of length $j$, membership vectors of the starting points are
traces of the reversed $j$-step system on its selected endpoints.
Theorem~\ref{up:layered} bounds their number by
$(s_\omega+1)^{h_j}$.
Multiplying over the fixed paths gives at most
\begin{equation}
 \prod_\omega(s_\omega+1)^{h_{|\omega|}}
 \le (s+1)^{c_\Delta h_\Delta}=(s+1)^{d_\Delta}.
 \label{alg:patterns}
\end{equation}
One range can be listed in $\widetilde O_\Delta(n)$ time.
Start with its endpoint and move backwards through its layers.
At each step, an orthogonal emptiness structure on the marked points finds
all points in the preceding layer with a marked successor.
There are at most $\Delta$ steps, each involving at most $n$ points.

\paragraph*{Propagation through the prefix tree.}
\label{alg:prefix}
Ignoring polylogarithmic factors, we balance $n/t$ and $t^{d_\Delta}$ in
Lemma~\ref{alg:ordering}.
This gives $t=\lceil n^\alpha\rceil$, since
$\alpha=1/(d_\Delta+1)$.
Sample from the list of labels of $\mathcal F_B$ and let $\lambda$ be the
resulting ordering of $X$.
Listing the samples and forming their membership vectors takes
$\widetilde O_\Delta(nt)$ time.
We represent every range for this starting cell in the same ordering.
The cell paths form a rooted tree under the prefix relation.
At the root, each vertex of $X$ has its singleton interval.
Suppose the ranges at a parent path $\omega'$ have been computed.
For each parent endpoint $u$, create a descriptor $(u,I)$ for each maximal
interval $I$ of $F_{\omega',u}$.
For a child path $\omega$ and every endpoint $v$ in its last cell,
apply Lemma~\ref{alg:interval} to compute
\begin{equation}
 F_{\omega,v}=\bigcup_{u\lhd v}F_{\omega',u}.
 \label{alg:recurrence}
\end{equation}
This determines all ranges by induction on the path length.

For each vertex $v$, merge the intervals from all length-$\Delta$ paths ending
in its cell.
Their union is exactly $N^\Delta[v]\cap X$: graph paths specify cell paths,
and shorter paths can be extended by staying at their endpoint.
Such stationary steps satisfy the weak comparisons within a cell.
Conversely, removing stationary steps from a layered chain gives a graph
walk with at most $\Delta$ edges.
Return no if this union differs from $[1,a]$ for some $v$.
Repeat for every nonempty starting cell, and return yes if all tests pass.
Every pair of vertices is tested in the iteration containing its first
vertex, so the answer is correct.

\paragraph*{Running time.}
Let $K$ count the maximal intervals in all ranges of $\mathcal F_B$.
Each path has at most $27$ children.
Thus the total number of input descriptors over all applications of
Lemma~\ref{alg:interval} is at most $27K$.
The maximal intervals of one range are disjoint and have total length equal
to the size of that range, which is at most $a$.
Copying them to at most $27$ children gives total input interval length at
most $27ma$.
There are at most $m$ queries, with at most $K$ output intervals in total.
For any fixed $b\ge1$, propagation and final merging therefore take
\[
 \widetilde O_\Delta\left((K+m)b+\frac{n^2}{b}\right)
\]
time.
Our choice of $t$ satisfies $n/t\le n^{1-\alpha}$ and
$(t+1)^{d_\Delta}=O_\Delta(n^{1-\alpha})$.
Lemma~\ref{alg:ordering} and~\eqref{alg:patterns} therefore give
$\mathbb E[K]=\widetilde O_\Delta(n^{2-\alpha})$.
The propagation bound is linear in $K$, so we may take its expectation for
this same sampled ordering.
We now balance $n^{2-\alpha}b$ and $n^2/b$, choosing
$b=\lceil n^{\alpha/2}\rceil$.
Including the sampling cost $\widetilde O_\Delta(nt)$, the expected time is
\[
 \widetilde O_\Delta
 \left(n^{1+\alpha}+n^{2-\alpha/2}\right)
 =\widetilde O_\Delta(n^{2-\alpha/2}),
\]
where the equality uses $\alpha\le1/2$.
There are only $O_\Delta(1)$ starting cells.
Absorbing the polylogarithmic factor into $n^{\alpha/4}$ proves
Theorem~\ref{thm:diameter}.
Sampling affects only the ordering and the time bound.
All propagated ranges are exact for every ordering, so the algorithm is
Las Vegas.

\section{A cubic lower bound}
\label{sec:lower}

For the lower bound in Theorem~\ref{thm:lower}, we construct a rooted tree whose leaves have the same depth.
We replace each tree vertex by a clique whose vertices carry subset labels.
Between adjacent cliques, an edge joins two vertices exactly when their labels agree.
At each leaf, we retain the labels that contain that leaf and add a target vertex.
Let $D$ be the common leaf depth, let $x_i$ be the target at leaf $i$, and let $w_S$ be the root vertex with subset label $S$.
The encoding will give the exact distances
\begin{equation}\label{low:main-distances}
 \operatorname{dist}(w_S,x_i)=
 \begin{cases}
 D+1,&i\in S,\\
 D+2,&i\notin S.
 \end{cases}
\end{equation}
Thus one fixed radius distinguishes membership in every subset of the targets.
This use of cliques, equal-label matchings, and coordinate bits appears in the
diameter lower bounds of~\cite{bringmann2022diameter}.
Here we place the cliques on a branching tree in three dimensions.
The tree has $8^h$ leaves at depth $O(2^h)$, which gives the cubic dependence on the radius.

\begin{figure}[tb]
\centering
\includegraphics[width=0.86\textwidth]{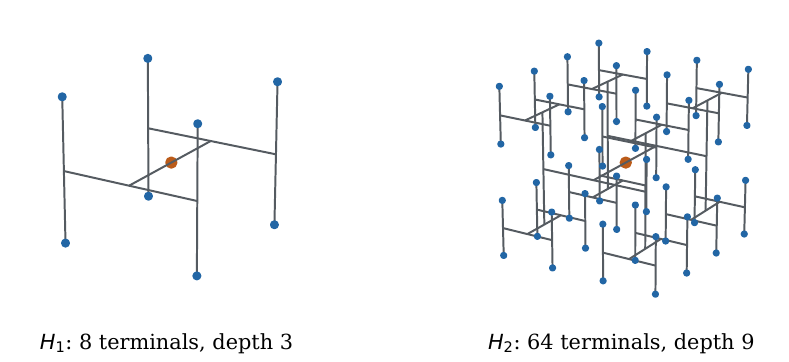}
\caption{The base trees $H_1$ and $H_2$ before subdivision, with $8$ and $64$ leaves at depths $3$ and $9$, respectively. Roots are orange and designated terminals are blue. All terminals of $H_h$ have depth $3(2^h-1)$. Label cliques and target vertices are not shown.}
\label{fig:grid-trees}
\end{figure}

\paragraph*{An equal-depth grid tree.}
We construct a tree $H_h$ from axis-parallel unit grid edges.
At scale $h$, a three-level branching structure joins the origin to eight translated
copies of the construction at scale $h-1$.
The copies lie in disjoint octants and meet the branching structure only at their roots.
Each of the three branching levels contributes $2^{h-1}$ edges to a root-to-copy path.
Consequently, the common depth satisfies $D'_0=0$ and $D'_h=3\cdot2^{h-1}+D'_{h-1}$.
Solving this recurrence gives $D'_h=3(2^h-1)$, while the number of leaves is $8^h$.
Figure~\ref{fig:grid-trees} shows the first two trees.
We double the coordinates and insert the midpoint of every edge.
Original vertices then have only even coordinates, and each midpoint has exactly one odd coordinate.
A pair at $\ell_1$ distance one must consist of a midpoint and one of its two endpoints.
The resulting grid representation is therefore induced, and every point has at most one odd coordinate.
A path of length $t$ attached at the root gives a tree of leaf depth
$D=6(2^h-1)+t$.
The recursive tree and these two modifications are proved in
Appendices~\ref{low:tree-construction} and~\ref{low:subdivision}.

\paragraph*{Realizing the labels with cubes.}
We apply the linear map
\[
 A(x,y,z)=(x+y-z,\ x-y+z,\ -x+y+z).
\]
On our grid points, two distinct images have $\ell_\infty$ distance at most~$1$
if and only if their preimages are adjacent in the tree.
The parity condition is essential here.
Indeed, an integer vector mapped by $A$ into $[-1,1]^3$ is either zero, a unit grid direction, or one of $\pm(1,1,1)$.
The last two vectors cannot be differences of points that each have at most one odd coordinate.
All nonadjacent pairs consequently have image distance at least two.
Appendix~\ref{low:linear-map-section} verifies these cases.

Put $k=8^h$ and $m=2^k$.
Number the subset labels by integers $0\le s<m$, with binary digits $s_i$, and write
$S(s)=\{i:s_i=1\}$.
Choose the rational parameters
\[
 Q=8(m+1),\qquad \lambda=1-\frac1Q,
 \qquad \varepsilon_s=\frac{2(s+1)}Q.
\]
At base point $b$, label $s$ has centre $p(b,s)=\lambda A(b)+\varepsilon_s\mathbf1$,
where $\mathbf1=(1,1,1)$.
Every base point other than a designated leaf carries all labels.
At leaf $b_i$, retain exactly the labels with $s_i=1$ and put $x_i=\lambda A(b_i)$.
If $o$ is the root, then $w_{S(s)}=p(o,s)$, denoted by $w_s$ in Appendix~\ref{ap:lower}.

All offsets lie in $[0,1/4)$, so centres at one base point form a clique.
For adjacent base points $b,c$, every coordinate of $A(b-c)$ is $1$ or $-1$, and both signs occur.
Writing $\delta$ for the difference of two offsets, we obtain
\[
 \|\lambda A(b-c)+\delta\mathbf1\|_\infty
 =\max\{|\lambda+\delta|,|{-\lambda}+\delta|\}
 =\lambda+|\delta|.
\]
Equal labels give distance $\lambda<1$.
Unequal offsets differ by at least $2/Q$, so they give distance at least $1+1/Q>1$.
For nonadjacent base points the distance is at least $2\lambda-2m/Q=7/4>1$.
Offset zero occurs only at designated leaves, whose coordinates are all even.
Two such base points cannot be adjacent, so no base point adjacent to a target's base point carries offset zero.
The neighbours of $x_i$ are therefore exactly the retained labels at its own leaf.
These statements classify all edges, including edges incident to targets;
Appendix~\ref{low:coordinates} gives the complete verification.

\paragraph*{Distances and the choice of radius.}
Project a path in the cube graph to its base points.
Each edge either traverses an edge of the base tree or leaves the base point unchanged.
Every path from $w_S$ to $x_i$ therefore uses at least $D$ edges between cliques
and one final edge inside the leaf clique.
When $i\in S$, following label $S$ along the tree and then taking the target edge gives length $D+1$.
A path of this length has no other edge inside a clique.
All its first $D$ edges preserve the label, so its penultimate vertex has label $S$ at leaf $i$.
That vertex is present exactly when $i\in S$.
If $i\notin S$, one change of label at the root gives a path of length $D+2$.
This proves~\eqref{low:main-distances}; see also Appendix~\ref{low:distance-section}.
Thus the targets are shattered at radius $D+1$.
For an arbitrary integer $r\ge7$, take
\[
 h=\left\lfloor\log_2\frac{r+5}{6}\right\rfloor,
 \qquad t=r-(6\cdot2^h-5).
\]
The choice of $h$ gives $h\ge1$ and $t\ge0$, and the added root path makes $D+1=r$.
Moreover, $2^h>(r+5)/12$, so cubing gives $8^h>(r+5)^3/1728$.
Every witness and target is present in this one rational construction and uses the same radius $r$.
Appendix~\ref{low:all-radii} gives the full parameter calculation for Theorem~\ref{thm:lower}.

\section{Conclusion}
We establish bounded VC-dimension for every fixed radius in three-dimensional unit-cube graphs.
Combined with the interval-representation approach, this yields truly subquadratic diameter decision algorithms for every fixed threshold.
The cubic lower bound shows that dependence on the radius is necessary.
A natural next question is whether the doubly exponential upper bound can be replaced by a polynomial one.

\label{main:end}
\clearpage
\bibliography{references}

\clearpage
\appendix
\section{Coordinate normalization}\label{ap:normalization}

We give the full proof of Lemma~\ref{up:normalization}.
All points in this argument retain their layer labels.

\begin{proof}[Proof of Lemma~\ref{up:normalization}]
\paragraph*{Perturbing coordinates.}
First fix one coordinate.
We assign a distinct integer offset $\delta_j$ to each layer $L_j$.
If the comparison from $L_{j-1}$ to $L_j$ is $<$ or $\ge$, require
$\delta_{j-1}>\delta_j$.
If it is $>$ or $\le$, require $\delta_{j-1}<\delta_j$.
A condition $\top$ imposes no constraint.
For each requirement, direct an edge from the layer with the smaller required offset to the layer with the larger required offset.
These requirements orient some edges of the path on layer indices $0,\ldots,t$.
The resulting directed graph is acyclic.
Assigning consecutive distinct integers in a topological order satisfies every requirement.

Choose distinct auxiliary numbers $\eta_p$ for all layer-labelled points.
For a point $p\in L_j$, replace its coordinate $p_d$ by
\[
 p_d+\epsilon\delta_j+\epsilon^2\eta_p,
\]
where $\epsilon>0$ is sufficiently small.
There are finitely many coordinates and prescribed strict orders.
We can therefore choose $\epsilon$ so that every comparison between originally unequal coordinates, and every prescribed strict order, retains its sign.
If two originally equal coordinates are in different layers, their new order is determined by the distinct offsets $\delta_j$.
If they are in the same layer, their new order is determined by the distinct numbers $\eta_p$.
Thus all coordinates become distinct.

Consider an originally equal pair in two consecutive layers.
For a $<$ condition, the left offset is larger, so the new strict comparison is false, as before.
For a $\ge$ condition, the left offset is larger, so replacing $\ge$ by $>$ gives a true comparison, as before.
For a $>$ condition, the left offset is smaller, and the new comparison remains false.
For a $\le$ condition, the left offset is smaller, so replacing $\le$ by $<$ gives a true comparison.
The coordinate perturbation therefore preserves all adjacent-layer conditions.
Perform it separately in each of the three coordinates.
After replacing the weak comparison symbols as described, the represented layered system has only strict comparisons and $\top$.

\paragraph*{Normalizing the first relation.}
Now suppose that $t\ge1$.
For each coordinate whose first relation is $\top$, translate all points in $L_1,\ldots,L_t$ together far enough in the positive direction that every point of $L_0$ has a smaller coordinate than every point of $L_1$.
We may choose the translation to avoid the finitely many values that would create coordinate equalities with $L_0$.
Replacing this first condition by $<$ preserves all its truth values.
Every later relation has both endpoints translated by the same amount, so its truth values are also preserved.
Finally, reflect any coordinate whose first comparison is $>$, simultaneously in all layers, and reverse the comparison symbol in that coordinate in every relation.
All first comparisons are now $<$.
The translations do not change the groups in $L_0$, and a reflection only reverses the common direction of an ordered band.
Thus both the represented relations and the ordered-band property are preserved.
\end{proof}

\section{Counting lemmas}\label{ap:count}

This appendix proves the trace bound in Lemma~\ref{up:trace-counting} using the Sauer--Shelah--Perles argument~\cite{sauer1972density,shelah1972stability}.
The three-family bound and the monotone subsequence lemma are proved in Section~\ref{sec:upper}.

\begin{proof}[Proof of the trace bound in Lemma~\ref{up:trace-counting}]
We first prove the bound $\sum_{j=0}^g\binom mj$.
It suffices to consider a family $\mathcal F$ of distinct subsets of an $m$-element ground set.
For $m=0$, the family contains at most one set.
For $g=0$, two distinct sets would shatter any point in their symmetric difference, so again the family contains at most one set.
These are the base cases.

For $m,g\ge1$, choose a ground-set element $x$ and define two families on the other $m-1$ elements:
\[
 \mathcal A=\{F\in\mathcal F:x\notin F\},
 \qquad
 \mathcal B=\{F\setminus\{x\}:F\in\mathcal F,\ x\in F\}.
\]
Their union is the family of all traces after deleting $x$, so its VC-dimension is at most $g$.
Their intersection has VC-dimension at most $g-1$.
To see this, suppose that it shattered a set $Y$ of size $g$.
Each trace on $Y$ would then be realized both by a set of $\mathcal F$ containing $x$ and by one excluding $x$.
Thus $\mathcal F$ would shatter $Y\cup\{x\}$, contrary to its VC-dimension bound.
By induction on $m$,
\begin{align*}
 |\mathcal F|
 &=|\mathcal A|+|\mathcal B|
  =|\mathcal A\cup\mathcal B|+|\mathcal A\cap\mathcal B|\\
 &\le\sum_{j=0}^g\binom{m-1}{j}
       +\sum_{j=0}^{g-1}\binom{m-1}{j}
  =\sum_{j=0}^g\binom mj.
\end{align*}
As usual, a binomial coefficient is zero when its lower index exceeds its upper index.
For the further bound, encode a subset of $[m]$ of size at most $g$ by its increasing list of elements, padded at the end with zeros to length $g$.
This is an injection into $\{0,1,\ldots,m\}^g$, giving at most $(m+1)^g$ subsets.

\end{proof}

\section{Explicit numerical bounds}\label{ap:numeric}

Section~\ref{up:grid} derives the general bound $f^*(r)\le2e_r^2-1$ and proves $f^*(0)=1$.
Here we verify the explicit radius-four estimates stated after Theorems~\ref{thm:vc} and~\ref{thm:diameter}.

For radius four,
\begin{equation}\label{ap:radius-four-constants}
 h_4=6^{120},\qquad
 e_4=531441\cdot6^{120}+12,
 \qquad
 d_4=551881\cdot6^{120}.
\end{equation}
Here $27^4=531441$, $\lceil\log_2(9^3)\rceil=10$, and
$1+27+27^2+27^3+27^4=551881$.
We verify the bounds $e_4<2^{330}$ and $d_4+1<2^{330}$ with integer arithmetic.
Let $A=531441\cdot6^{120}=2^{120}3^{132}$.
Since
\[
 3^{22}=31381059609<32212254720=15\cdot2^{31}
       =\frac{15}{16}2^{35},
\]
raising to the sixth power and multiplying by $2^{120}$ gives
$A<(15/16)^6 2^{330}$.
For every positive integer $j$, $(15/16)^j\le15/16$.
Therefore
\[
 A+12<\frac{15}{16}2^{330}+12<2^{330},
\]
where the last inequality follows from $12<2^{326}$.
Also,
\[
 15\cdot551881=8278215<8503056=16\cdot531441.
\]
It follows that
\[
 d_4+1<\frac{16}{15}A+1
       <\left(\frac{15}{16}\right)^5 2^{330}+1
       \le\frac{15}{16}2^{330}+1
       <2^{330}.
\]

Now put $k_0=2^{339}$.
Since $k_0+1<2^{340}$, we have
\[
 e_4\log_2(k_0+1)<340\cdot2^{330}<512\cdot2^{330}=k_0.
\]
The trace bound is strictly smaller than $2^{k_0}$, so no set of $k_0$ vertices is shattered.
This proves the bound $2^{339}-1$.
Finally, the inequality $d_4+1<2^{330}$ gives
\[
 \frac{1}{4(d_4+1)}>2^{-332},
\]
which yields the stated explicit exponent in Theorem~\ref{thm:diameter}.

\section{Data structures and sampling for the diameter algorithm}
\label{ap:algorithm}

We give the data structures used in Section~\ref{sec:algorithm} and prove
Lemmas~\ref{alg:interval} and~\ref{alg:ordering}.
The interval coverage and coloured-query approach follows~\cite{chan2025vc}.
We give complete proofs for the boundary conventions and labelled ranges
needed in our application.
We first construct orthogonal and coloured dominance structures, then use
them for interval propagation in Appendix~\ref{ap:interval-proof}.
Appendix~\ref{ap:ordering-proof} proves the sampled-ordering bound.
Appendix~\ref{ap:cell-path-accounting} verifies the implementation and accounts
for all intermediate ranges.
Throughout this appendix, interval lengths count integer positions.
All logarithms in running-time bounds may be replaced by
$\log(N+a+n+2)$.
Thus empty sets and sets of size one cause no exceptional logarithmic terms.

\subsection{Orthogonal counting}

We use the standard recursive range-tree construction
(see~\cite[Chapter~5]{deberg2008computational}).
The proof below records its treatment of strict bounds and multiplicities.

\begin{lemma}
\label{ap:orthogonal}
For fixed $D$, a multiset of $N$ points in $\R^D$ can be preprocessed in
$\widetilde O(N+1)$ time and space to support orthogonal prefix counting in
$O(\log^D(N+2))$ time.
Each coordinate bound may be strict or weak.
The same bounds hold for emptiness queries.
\end{lemma}

\begin{proof}
In one dimension, sort the points and store the number of points in each
prefix, counting multiplicities.
Binary search finds the last coordinate below a strict threshold, or the
last coordinate at most a weak threshold.
The prefix count is then available in constant additional time.

For $D>1$, build a balanced binary tree on the first coordinates.
At each node, recursively store a $(D-1)$-dimensional structure for the points
below that node.
The first-coordinate prefix of a query decomposes into
$O(\log(N+2))$ disjoint node sets.
Query their associated structures on the other coordinates and add the
counts.
Induction gives query time $O(\log^D(N+2))$.

At each tree depth, the node sets are disjoint and contain $N$ points in total.
There are $O(\log(N+2))$ depths.
Induction on $D$ therefore bounds both the space and construction time by
$N$ times a fixed power of $\log(N+2)$, with a constant-size representation
when $N=0$.
Testing whether the resulting count is zero gives an emptiness query.
\end{proof}

Lower bounds on a coordinate are changed into upper bounds by reflection.
Consequently the lemma also supports any conjunction of fixed coordinate
comparisons.
In particular, it supports the generalized dominance queries used below.

\subsection{Coloured dominance in three dimensions}

\begin{lemma}
\label{ap:coloured}
Let $N$ points of $\R^3$ have colours from a prescribed set of $C$ required
colours.
After $\widetilde O(N+1)$ preprocessing, a query point can be tested in
$\widetilde O(1)$ time for whether it dominates a point of every required
colour.
Only the value of $C$ and the colours present in the input need to be stored.
\end{lemma}

\begin{proof}
A point $(a,b,c)$ is dominated by a query $(x,y,z)$ precisely when the query
belongs to
\[
 [a,\infty)\times[b,\infty)\times[c,\infty).
\]
We first process the points of one colour.
Sort them by $a$, breaking ties in any fixed order.
In the $yz$-plane, maintain the union of the quadrants
$[b,\infty)\times[c,\infty)$ already inserted.
This union has the form
\[
 \{(y,z):z\ge f(y)\},
\]
where $f$ is a nonincreasing step function and may take the value $+\infty$.
Store its finite breakpoints in a balanced search tree ordered by $y$.
After redundant breakpoints are removed, their heights strictly decrease.
At a breakpoint the new, lower height applies.

Consider the insertion of $(a,b,c)$.
If $f(b)\le c$, monotonicity gives $f(y)\le c$ for all $y\ge b$, so the
new quadrant adds no points.
Otherwise, its new part is
\begin{equation}
 \{(y,z): y\ge b,\ c\le z<f(y)\}.
 \label{ap:new-region}
\end{equation}
Traverse the old staircase from $b$ until its height first becomes at most
$c$.
For each old constant-height interval met by this traversal, output the
nonempty rectangle
\[
 [\ell,u)\times[c,h),
\]
where $h$ is the old height and $[\ell,u)$ is the traversed part of that
interval.
The last interval may be unbounded.
These rectangles partition~\eqref{ap:new-region} exactly.
The left endpoint is included and the right endpoint is excluded, as
required by the definition of the staircase.

The update replaces the traversed part of the staircase by height $c$.
It introduces at most one new breakpoint.
Apart from at most a constant number of rectangles at the ends, every
rectangle can be charged to a breakpoint deleted by this update.
Each input point introduces at most one breakpoint, and each breakpoint
is deleted at most once.
The total number of rectangles is therefore linear in the number of points
of this colour.
Predecessor searches, insertions, and deletions take $O(\log(N+2))$ time
each, so the entire sweep takes $O((N+1)\log(N+2))$ time over all colours.

Extend each rectangle created at sweep coordinate $a$ to a box
\begin{equation}
 [a,\infty)\times[\ell,u)\times[c,h).
 \label{ap:colour-box}
\end{equation}
The $yz$-projections of boxes of one colour are pairwise disjoint.
Indeed, a rectangle is created only for points outside the old union, and
the union only increases during the sweep.
For a fixed $(y,z)$ covered by some quadrant, the unique rectangle containing
it was created when that point first became covered.
Its sweep coordinate is the smallest $a$ of an input point with
$b\le y$ and $c\le z$.
It follows that the boxes of this colour have exactly the same union as its
three-dimensional upper orthants.
They are also pairwise disjoint.
Ties in $a$ do not affect this conclusion, because all boxes created at that
coordinate include their first-coordinate boundary.

Construct these boxes for every colour.
There are $O(N)$ boxes in total.
A query lies in at most one box of each colour, and it lies in such a box
exactly when it dominates an input point of that colour.
Thus the number of boxes containing the query is the number of colours hit.

Represent a box~\eqref{ap:colour-box} by the five-dimensional point
\[
 (a,\ell,-u,c,-h).
\]
The query $(x,y,z)$ belongs to the box exactly when
\[
 a\le x,\qquad \ell\le y,\qquad -u<-y,
 \qquad c\le z,\qquad -h<-z.
\]
Lemma~\ref{ap:orthogonal} counts these points with a mixed strict and weak
prefix query.
Infinite endpoints can be represented by ordered sentinels, since query
coordinates are finite.
Compare the count with $C$.
If a required colour has no input point, it contributes no box and the count
cannot reach $C$.
If $C=0$, the answer is yes.
This proves both the claimed query time and the treatment of absent colours.
\end{proof}

For a generalized dominance relation, reflect coordinates that have the
opposite direction.
A coordinate with the predicate $\top$ can be replaced by zero in every
input and query point.
The comparisons between neighbouring grid cells are weak, so this gives
exactly the queries required by the diameter algorithm.
Strict comparisons in the auxiliary counting structures are handled by
Lemma~\ref{ap:orthogonal}.

\subsection{Covering, avoiding, and extracting integer intervals}
\label{ap:interval-proof}

We prove Lemma~\ref{alg:interval}.
Assume $a\ge1$, since for $a=0$ every output is empty.
Reflect coordinates so that all nontrivial conditions in $p\lhd q$ are
upper dominance conditions.
Let the input descriptors be $(p,I_p)$, and let
\[
 U(q)=\bigcup_{p\lhd q}I_p.
\]
The same point may occur in several descriptors.
Each occurrence keeps its own interval.

\paragraph*{Block structures.}
Partition $[1,a]$ into consecutive blocks of length $b$, with a possibly
shorter last block.
For every block fully contained in $I_p$, place one copy of $p$ in the
block's full-coverage structure.
A dominance emptiness query in this structure tests whether one selected
input interval covers the entire block.
An input interval contains at most $|I_p|/b+1$ full blocks, where the extra
one accounts for the last, shorter block.
The total number of copies is therefore $O(N+L/b)$.

Every input interval meets at most two further blocks without containing
them.
For each such partial block and each integer position $i$ in its intersection
with $I_p$, insert a copy of $p$ with colour $i$ into that block's partial
coverage structure.
The required colours are all positions of the block.
Also insert $p$ into a single-colour structure associated with position $i$.
There are at most $2Nb$ copies of each kind.
Build the block structures by Lemma~\ref{ap:coloured}, and the single-colour
structures by Lemma~\ref{ap:orthogonal}.
Only nonempty structures are stored.
A lookup for a missing structure returns the empty answer.
The required number of colours of a block is its length, even if some of its
positions never occur in the partial input.
Thus missing colours are counted correctly without allocating a structure
for every position.
The total preprocessing time and space are
\begin{equation}
 \widetilde O\left(Nb+\frac Lb+1\right).
 \label{ap:block-preprocessing}
\end{equation}

\paragraph*{Coverage queries.}
For an integer interval $I$, we test whether $I\subseteq U(q)$.
Visit the blocks meeting $I$.
In one such block, first ask whether a selected full-coverage descriptor
exists.
If it does, all positions of $I$ in this block are covered.
If it does not and the whole block belongs to $I$, query the block's partial
coverage structure for all its required colours.
If it does not and only part of the block belongs to $I$, query the
single-colour structure for each position of $I$ in that block.
These tests are exact: each descriptor meeting a block either covers it
fully or was included among its partial descriptors.

There are at most two partially queried blocks, contributing at most $2b$
single-position queries.
There are $O(1+|I|/b)$ blocks in total.
The coverage query therefore takes
\begin{equation}
 \widetilde O\left(b+\frac{|I|}{b}\right)
 \label{ap:coverage-query}
\end{equation}
time, including when $b>a$.

\paragraph*{Avoidance queries.}
For $I=[l,r]$, we test whether $I\cap U(q)=\varnothing$.
An input interval $I_p=[l_p,r_p]$ intersects $I$ exactly when
$l_p\le r$ and $r_p\ge l$.
Together with the three coordinate comparisons of $p\lhd q$, these are
five dominance conditions on $(p,l_p,-r_p)$.
Lemma~\ref{ap:orthogonal} gives an emptiness structure with
$\widetilde O(N+1)$ preprocessing and $\widetilde O(1)$ query time.
It answers the avoidance query exactly.

\paragraph*{Extraction.}
Start with $[1,a]$.
For a current interval $I$, output it if it is covered by $U(q)$.
Discard it if it avoids $U(q)$.
Otherwise, split it into two consecutive intervals of nearly equal length
and recurse.
A singleton is either covered or avoided, so this process terminates.
The output consists of disjoint covered intervals whose union is $U(q)$.
Sort them by their left endpoints and merge adjacent intervals to obtain
its maximal intervals.

Let $k_q$ be the number of these maximal intervals.
A node at which the recursion continues contains both a covered and an
uncovered integer position.
It therefore contains a transition between consecutive positions with
different membership in $U(q)$.
There are at most $2k_q$ such transitions.
At any fixed depth, the current intervals are disjoint, so at most $2k_q$
nodes continue the recursion.
Every queried nonroot node is a child of a continuing node.
There are consequently
\[
 O(1+k_q\log(a+1))
\]
queries altogether, including the cases $U(q)=\varnothing$ and
$U(q)=[1,a]$.
At one depth, the queried intervals have total length at most $a$.
Over all depths their total length is $O(a\log(a+1))$.
Summing~\eqref{ap:coverage-query} over these intervals gives extraction time
\[
 \widetilde O\left((1+k_q)b+\frac ab\right).
\]
Avoidance queries and the final merging fit in this bound.
Summing over the $m$ query points, and adding
\eqref{ap:block-preprocessing}, gives
\[
 \widetilde O\left((N+m+K')b+\frac{L+ma}{b}+1\right),
\]
where $K'=\sum_q k_q$.
This proves Lemma~\ref{alg:interval}.

\subsection{A common random ordering}
\label{ap:ordering-proof}

We now prove Lemma~\ref{alg:ordering}.
If $m=0$, choose any fixed ordering and return it.
There are no represented ranges, so $K=0$.
The same conclusion holds if $X$ is empty.
Assume henceforth that $m\ge1$ and $1\le a=|X|\le n$.

Independently sample $t$ range labels, uniformly with replacement from
$\mathcal F$.
List each sampled range and record the $t$-bit membership vector of each
element of $X$.
Group equal vectors and place every group consecutively in the ordering.
Use any fixed order within and between the groups.
The number of groups is at most $(t+1)^d$ by hypothesis.
Repeated samples do not add distinct membership patterns, so the hypothesis
also applies when sampled labels repeat.
Listing the ranges costs $\widetilde O(nt)$.
Recording and sorting the vectors takes $\widetilde O(at)$ additional time,
which is covered by the same bound.

For $x,y\in X$, let $c(x,y)$ be the number of range labels whose range
contains exactly one of $x,y$.
The probability that the two sampled membership vectors are equal is
\[
 \left(1-\frac{c(x,y)}m\right)^t
 \le \exp\left(-\frac{t c(x,y)}m\right).
\]
If $c(x,y)>6m\log(n+1)/t$, this probability is at most $(n+1)^{-6}$.
There are at most $n^2$ pairs.
With probability at least $1-(n+1)^{-4}$, every pair in one sampled group
therefore satisfies
\begin{equation}
 c(x,y)\le \frac{6m\log(n+1)}t.
 \label{ap:good-sample}
\end{equation}
This estimate also holds when the threshold exceeds $m$, since then there
is no pair violating it.

Call two consecutive elements of the final ordering a crossing of a range
if exactly one of them belongs to that range.
On the event~\eqref{ap:good-sample}, the at most $a-1$ adjacencies inside
sampled groups contribute at most
$6am\log(n+1)/t$ crossings in total over all range labels.
There are at most $(t+1)^d-1$ adjacencies between groups, each contributing
at most $m$ crossings.
The number of maximal intervals in one range is at most its number of
crossings plus one.
Hence, on this event,
\[
 K\le m\left(1+\frac{6n\log(n+1)}t+(t+1)^d\right).
\]
For every sample, we also have $K\le ma\le mn$.
The contribution of the complementary event to the expectation is at most
$mn(n+1)^{-4}\le m$.
Combining the two estimates proves
\[
 \mathbb E[K]
 =\widetilde O\left(m\left(1+\frac nt+(t+1)^d\right)\right).
\]
This completes the proof of Lemma~\ref{alg:ordering}.

\subsection{Implementing and accounting for the cell-path computation}
\label{ap:cell-path-accounting}

We spell out how the preceding lemmas apply to the algorithm in
Section~\ref{sec:algorithm}.
After the coordinate-spread test, all nonempty cells have indices in
$\{0,\ldots,\Delta\}^3$ relative to the chosen origin.
The comparisons with the grid boundaries use the half-open convention,
including points whose coordinate is exactly an integer offset.
No floor operation on unbounded real numbers is needed.
For neighbouring cells, their local coordinates give the weak generalized
dominance relation derived in Section~\ref{up:grid}.
These local coordinates are used as geometric objects in the interval
structures.

\paragraph*{Labels and sampling.}
For the fixed starting cell $B$, form a list containing the pair
$(\omega,v)$ for each cell path of length at most $\Delta$ and each vertex
in its last cell.
This list has length $m\le c_\Delta n$ and can be generated in
$O_\Delta(n)$ time.
The root path contributes one label for every vertex of $X$.
In particular, $m\ge a\ge1$ in this application.
Two labels are retained even if their ranges happen to be identical.
Uniform sampling in Lemma~\ref{alg:ordering} is from this list of labels.

For selected labels belonging to one path, repeated endpoint labels impose
repeated membership coordinates.
Deleting these repetitions cannot increase the number of membership
vectors.
The reversed-system trace bound therefore gives
$(s_\omega+1)^{h_{|\omega|}}$ whether or not selected endpoints repeat.
Different paths remain different groups in the product
\eqref{alg:patterns}.
This proves the required dual pattern bound for the full labelled family.

\paragraph*{Listing one sampled range.}
Let $\omega=(B_0,\ldots,B_j)$, with $B_0=B$, and let its endpoint be
$v\in P\cap B_j$.
Mark just $v$ in layer $j$.
For $i=j,j-1,\ldots,1$, build an orthogonal emptiness structure on the marked
points of layer $i$.
For each point $u$ in layer $i-1$, mark $u$ if there is a marked point $w$
with $u\lhd_i w$.
For a query $u$, these are reflected orthogonal comparisons on the stored
points $w$, so Lemma~\ref{ap:orthogonal} applies.
An empty marked layer yields empty marks in every preceding layer.

By backward induction, a point marked in layer $i$ has a path to $v$ through
the remaining prescribed cells, and every such point is marked.
The final marked set is therefore $F_{\omega,v}$.
Each layer has at most $n$ points, and there are at most $\Delta$ layers.
The total cost is $\widetilde O_\Delta(n)$.
For $j=0$, simply return the singleton $\{v\}$.
Listing all $t$ sampled ranges thus costs $\widetilde O_\Delta(nt)$.

\paragraph*{Exact propagation.}
Process all cell paths in increasing order of length.
At the root, the range of $u\in X$ is the singleton containing its position
in $\lambda$.
Suppose every parent range has already been computed exactly.
A starting point reaches a child endpoint $v$ precisely when it reaches
some parent endpoint $u$ with $u\lhd v$.
Replacing each parent range by its maximal intervals gives exactly the
descriptor union in Lemma~\ref{alg:interval}.
Its output is therefore the child range by~\eqref{alg:recurrence}.
Induction proves exactness at every path node.
Repeated visits to a physical cell use its points as a fresh layer with
the same coordinates.
The weak comparisons permit a physical vertex to remain in place between
consecutive equal cells.

\paragraph*{Total input size.}
Let $k_{\omega,v}$ be the number of maximal intervals of $F_{\omega,v}$.
Then
\[
 K=\sum_{(\omega,v)}k_{\omega,v},
 \qquad
 \sum_{(\omega,v)}|F_{\omega,v}|\le ma.
\]
A parent path has at most $27$ children.
Its endpoint intervals are copied once for each child, so the sum of all
descriptor counts is at most $27K$.
Maximal intervals of a single range are disjoint.
Their lengths therefore sum to that range's cardinality, and the sum of all
input interval lengths is at most $27ma$.
Every nonroot label is queried only at its own path node.
Thus the total number of queries is at most $m$, and the total number of
output intervals is at most $K$.
There are at most $c_\Delta$ nodes, including empty paths.
The constant additive cost in Lemma~\ref{alg:interval}, summed over them,
is $O_\Delta(1)$.

Adding its bounds over all nodes yields
\[
 \widetilde O\left((27K+m+K)b+
       \frac{27ma+ma}{b}+c_\Delta\right)
 =\widetilde O_\Delta\left((K+m)b+\frac{n^2}{b}\right).
\]
For the last step, $a\le n$, $m\le c_\Delta n$, and $m\ge1$ absorb the
constant term.
For a fixed vertex $v$, concatenate the interval lists of the length-$\Delta$
paths ending at $v$.
Sorting by endpoints and merging adjacent or overlapping intervals takes
$\widetilde O(K+m)$ time over all vertices.
Vertices with no such label have the empty union.
Checking whether the union is $[1,a]$ then takes constant time per vertex.

\paragraph*{Expected time.}
The parameters of Section~\ref{sec:algorithm} satisfy
\[
 t=\lceil n^\alpha\rceil,\qquad
 b=\lceil n^{\alpha/2}\rceil,\qquad
 \alpha=\frac1{d_\Delta+1}.
\]
For $n\ge2$, we have
\[
 t+1\le3n^\alpha,
 \qquad (t+1)^{d_\Delta}
 \le3^{d_\Delta}n^{\alpha d_\Delta}
 =3^{d_\Delta}n^{1-\alpha}.
\]
Also $n/t\le n^{1-\alpha}$.
Lemma~\ref{alg:ordering}, with $m=O_\Delta(n)$, now gives
$\mathbb E[K]=\widetilde O_\Delta(n^{2-\alpha})$.
The bound for propagation is linear in $K$ once $b$ is fixed, so taking its
expectation is valid for this same sampled ordering.
Sampling and propagation together take expected time
\[
 \widetilde O_\Delta\left(
       n^{1+\alpha}+n^{2-\alpha}n^{\alpha/2}
       +n^{2-\alpha/2}\right)
 =\widetilde O_\Delta(n^{2-\alpha/2}).
\]
Here $1+\alpha\le2-\alpha/2$ follows from $\alpha\le1/2$.
For each fixed $\Delta$, the suppressed factor is a fixed power of
$\log(n+2)$.
It is $O_\Delta(n^{\alpha/4})$, which gives the stated
$O_\Delta(n^{2-1/(4(d_\Delta+1))})$ bound.

Finally, the sampled ordering changes only the representation lengths.
For every sample, each range query and each interval union is exact.
The completed length-$\Delta$ ranges cover every graph path of at most
$\Delta$ edges, including a path padded by stationary steps.
Conversely, every layered chain gives a graph walk after stationary steps
are removed.
The final answer is therefore correct for every sample.
The expectation concerns running time alone, as required for a Las Vegas
algorithm.

\section{The cubic lower-bound construction}
\label{ap:lower}

We give the full construction for Theorem~\ref{thm:lower}.
For integers $h\ge1$, $t\ge0$, and $1\le k\le8^h$, we construct a finite set
$P_{h,t,k}\subset\mathbb{Q}^3$.
It contains targets $x_1,\ldots,x_k$ and witnesses
$w_0,\ldots,w_{2^k-1}$ such that, for
\[
 D=6(2^h-1)+t,
\]
the distance from $w_s$ to $x_i$ is $D+1$ if the $(i-1)$st bit of $s$ is~$1$,
and $D+2$ if that bit is~$0$.
All distances are in the unit-cube graph of the constructed point set.

\subsection{An equal-depth tree in the integer grid}
\label{low:tree-construction}

We first construct a geometric tree $H_h$.
Its vertices are integer points on specified axis-parallel segments.
Its edges join consecutive integer points along those segments.
At this stage, an edge is included only when it belongs to a specified segment.

Let $H_0$ consist of the origin, which is both its root and its single designated terminal.
For $h\ge1$, put $a=2^{h-1}$ and take the following segments:
\begin{align*}
 &[-a,a]\times\{0\}\times\{0\},\\
 &\{\sigma a\}\times[-a,a]\times\{0\}
     &&\text{for }\sigma\in\{-1,1\},\\
 &\{\sigma a\}\times\{\tau a\}\times[-a,a]
     &&\text{for }\sigma,\tau\in\{-1,1\}.
\end{align*}
We call their union the skeleton.
For each $\sigma,\tau,\upsilon\in\{-1,1\}$, set
$c=(\sigma a,\tau a,\upsilon a)$ and attach a translated copy $c+H_{h-1}$.
The root of $H_h$ is the origin.
Its designated terminals are the designated terminals in the eight translated copies.

\begin{lemma}
\label{low:grid-tree}
For every integer $h\ge0$, the following properties hold.
\begin{enumerate}
\item The geometric realization of $H_h$ lies in
$[-(2^h-1),2^h-1]^3$.
\item Its only point in the plane $x=0$ is the root.
\item The segments meet only at the prescribed attachment points, and $H_h$ is a tree.
\item It has $8^h$ designated terminals, each at distance $3(2^h-1)$ from the root.
For $h\ge1$, these terminals are exactly its leaves.
\end{enumerate}
\end{lemma}

\begin{proof}
All four properties hold for $h=0$.
Suppose that they hold for $h-1$, and write $a=2^{h-1}$.
By the first induction hypothesis, the copy rooted at
$c=(\sigma a,\tau a,\upsilon a)$ lies in the box
\[
 B_c=c+[-(a-1),a-1]^3.
\]
Every coordinate of a point in $B_c$ has the sign of the corresponding coordinate of $c$,
and its absolute value is at least~$1$.
If $c$ and $c'$ have different sign triples, some coordinate of $B_c$ lies in
$(-\infty,-1]$ while that coordinate of $B_{c'}$ lies in $[1,\infty)$.
Thus the eight boxes are pairwise disjoint.

The $x$- and $y$-segments of the skeleton lie in the plane $z=0$.
They do not meet any box $B_c$.
A $z$-segment at $(x,y)=(\sigma a,\tau a)$ can meet only the two boxes whose
first two signs are $\sigma$ and $\tau$.
In the local coordinates of either translated copy, this segment lies on $x=y=0$.
The second induction hypothesis says that the copy meets the plane $x=0$ only at its root.
Consequently, the segment meets that copy only at the attachment point $c$.
This accounts for all possible intersections between the skeleton and the copies.

The skeleton itself is a tree.
Start with its $x$-segment and attach a $y$-segment at each endpoint.
These two $y$-segments are disjoint and each meets the $x$-segment only at its attachment point.
Next attach a $z$-segment at each of the four endpoints of the $y$-segments.
These four segments have distinct pairs of $x,y$ coordinates.
Each meets the earlier segments only at its attachment point.
Every attachment therefore preserves connectedness and creates no cycle.
The eight copies of $H_{h-1}$ are trees by induction.
Each shares exactly its root with the skeleton and is disjoint from the other copies.
Attaching them preserves connectedness and cannot create a cycle.
This proves the third property.

The path from the origin to any attachment point $c$ traverses an $x$-segment,
a $y$-segment, and a $z$-segment, using $a$ edges in each coordinate direction.
Its length is $3a$.
The distance from $c$ to a designated terminal in its copy is $3(a-1)$ by induction.
Hence every designated terminal has depth
\[
 3a+3(a-1)=3(2^h-1).
\]
The copies are disjoint, so their terminal sets are disjoint and contain
$8\cdot8^{h-1}=8^h$ terminals in total.
If $h=1$, the eight attachment points are precisely the leaves of the skeleton.
If $h>1$, the root of every attached copy has degree at least two within the copy.
After attachment it is not a leaf.
All other vertices retain their degrees, and every leaf of the copy is a designated terminal.
The skeleton has no remaining leaf outside the copies.
This proves the fourth property.

The skeleton lies in $[-a,a]^3$.
Each box $B_c$ lies in $[-(2a-1),2a-1]^3$, which proves the first property.
Every translated copy has $|x|\ge1$.
Every $y$- or $z$-segment of the skeleton has $|x|=a\ge1$.
The $x$-segment meets the plane $x=0$ only at the origin.
This proves the second property and completes the induction.
\end{proof}

\subsection{Subdivision and a path at the root}
\label{low:subdivision}

We next turn the chosen grid edges into all grid adjacencies among the constructed points.
For a finite subgraph $H=(V,E)$ of the three-dimensional integer grid, define
\[
 B(H)=\{2v:v\in V\}\ \cup\ \{u+v:\{u,v\}\in E\}.
\]
Thus we double every vertex coordinate and insert the midpoint of each doubled edge.

\begin{lemma}
\label{low:induced-subdivision}
Join two points of $B(H)$ when their $\ell_1$ distance is~$1$.
The resulting graph is exactly the graph obtained by subdividing each edge of $H$ once.
Every point of $B(H)$ has at most one odd coordinate.
\end{lemma}

\begin{proof}
Every original point $2v$ has three even coordinates.
If $\{u,v\}$ is an edge in coordinate direction $j$, its midpoint $u+v$ has an odd
$j$th coordinate and two even coordinates.
In particular, a midpoint is distinct from every original point.
Its odd coordinate determines the direction of the edge, and the two neighbouring even points
in that direction determine the endpoints.
Different edges therefore have different midpoints.

Two original points cannot have $\ell_1$ distance~$1$, since each coordinate difference is even.
Consider two midpoints.
If their odd coordinates have the same position, each coordinate difference is even,
so their $\ell_1$ distance is not~$1$.
If the odd coordinates have different positions, at least two coordinate differences are odd.
Their $\ell_1$ distance is then at least~$2$.

It remains to consider a midpoint $w=u+v$ and an original point $2z$.
Suppose that their $\ell_1$ distance is~$1$.
The only changed coordinate must be the odd coordinate of $w$.
Increasing or decreasing that coordinate by~$1$ gives exactly $2u$ or $2v$.
Thus a midpoint is adjacent only to the two endpoints of its own doubled edge.
Conversely, its distance to each of these endpoints is~$1$.
These cases include every pair of distinct points in $B(H)$, proving the claim.
\end{proof}

Apply Lemma~\ref{low:induced-subdivision} to $H_h$, and write $B_h=B(H_h)$.
With adjacency defined by $\ell_1$ distance~$1$, this is a tree whose designated terminals
are twice those of $H_h$.
Their common depth is
\[
 D_h=6(2^h-1).
\]
For an integer $t\ge0$, define
\[
 B_{h,t}=B_h\cup\{(0,0,-j):1\le j\le t\},
 \qquad o=(0,0,-t).
\]
We use $o$ as the new root and keep the designated terminals from $B_h$.

\begin{lemma}
\label{low:stem}
For $h\ge1$, the graph on $B_{h,t}$ with adjacency defined by $\ell_1$ distance~$1$
is obtained by attaching a path of length $t$ to the root of $B_h$.
Each designated terminal is at distance $D_h+t$ from the new root $o$.
Every point of $B_{h,t}$ has at most one odd coordinate.
\end{lemma}

\begin{proof}
The only segments of $H_h$ that have a point with $|x|<1$ belong to its root $x$-segment.
All other segments have $|x|\ge1$, by the construction and
Lemma~\ref{low:grid-tree}.
After doubling and subdivision, the only point of $B_h$ with $x=0$ is the origin,
and its only points with $|x|=1$ are $(1,0,0)$ and $(-1,0,0)$.
Every remaining point of $B_h$ has $|x|\ge2$.

For $j\ge1$, the new point $(0,0,-j)$ has $\ell_1$ distance $1+j\ge2$
from each of $(1,0,0)$ and $(-1,0,0)$.
Its distance from any old point with $|x|\ge2$ is at least~$2$.
Its distance from the origin is $j$, so it is adjacent to the origin exactly when $j=1$.
Two new points $(0,0,-j)$ and $(0,0,-j')$ are adjacent exactly when $|j-j'|=1$.
Thus the new points form a path that meets the old tree only through the edge to the origin.
The depth of every designated terminal increases by~$t$.
Only the third coordinate of a new point can be odd, proving the final assertion.
\end{proof}

For the remainder of the construction, the \emph{base tree} is the graph on $B_{h,t}$
from Lemma~\ref{low:stem}.
In particular, its edges are precisely the pairs at $\ell_1$ distance~$1$.
Its designated terminals are leaves, and the new root is distinct from all of them.

\subsection{From grid adjacency to cube adjacency}
\label{low:linear-map-section}

Define the invertible linear map
\begin{equation}
\label{low:linear-map}
 A(x,y,z)=(x+y-z,\ x-y+z,\ -x+y+z).
\end{equation}
Its inverse is
\[
 A^{-1}(u,v,w)=\left(\frac{u+v}{2},\frac{u+w}{2},\frac{v+w}{2}\right).
\]
Let $e_1,e_2,e_3$ be the standard basis vectors in $\mathbb{R}^3$.

\begin{lemma}
\label{low:parity-map}
Let $b,c\in\mathbb{Z}^3$ be distinct points, each with at most one odd coordinate.
Then
\[
 \|A(b-c)\|_\infty\le1
 \quad\Longleftrightarrow\quad
 b-c\in\{\pm e_1,\pm e_2,\pm e_3\}.
\]
All remaining pairs satisfy $\|A(b-c)\|_\infty\ge2$.
\end{lemma}

\begin{proof}
Suppose that $A(b-c)=(u,v,w)$ has $\ell_\infty$ norm at most~$1$.
Each of $u,v,w$ is an integer in $\{-1,0,1\}$.
Since $A^{-1}(u,v,w)$ is an integer vector, the sums $u+v$, $u+w$, and $v+w$
are all even.
Hence $u,v,w$ have the same parity.

If they are all even, then $u=v=w=0$ and $b=c$, contrary to the hypothesis.
Otherwise, each is either $1$ or $-1$.
Four of the eight cases are
\[
\begin{array}{c|c}
(u,v,w)&A^{-1}(u,v,w)\\ \hline
(1,1,1)&(1,1,1)\\
(1,1,-1)&e_1\\
(1,-1,1)&e_2\\
(-1,1,1)&e_3
\end{array}
\]
and the other four are their negatives.
The cases $b-c=\pm(1,1,1)$ would require $b$ and $c$ to have different parity
in all three coordinates.
This is impossible: the set of coordinates where their parities differ is the symmetric
difference of their sets of odd coordinates, and has size at most~$2$.
The six unit grid directions are the only remaining possibilities.

Conversely,
\[
 Ae_1=(1,1,-1),\qquad
 Ae_2=(1,-1,1),\qquad
 Ae_3=(-1,1,1),
\]
and these vectors and their negatives all have $\ell_\infty$ norm~$1$.
Finally, $A(b-c)$ is an integer vector.
If its norm is greater than~$1$, the norm is at least~$2$.
\end{proof}

Each image of a unit grid direction contains both a positive and a negative coordinate.
We use this fact to realize the equal-label matchings between neighbouring base points.

\subsection{The rational coordinates}
\label{low:coordinates}

Fix $h\ge1$, $t\ge0$, and $1\le k\le8^h$.
Choose any $k$ designated terminals of $B_{h,t}$ and call them $b_1,\ldots,b_k$.
Let
\begin{equation}
\label{low:parameters}
 m=2^k,\qquad Q=8(m+1),\qquad
 \lambda=1-\frac1Q,\qquad
 \varepsilon_s=\frac{2(s+1)}Q\quad(0\le s<m).
\end{equation}
The label $s$ represents the subset of $[k]$ given by its $k$ binary digits.
Write $s_i\in\{0,1\}$ for its $(i-1)$st bit.
Let $\mathbf{1}=(1,1,1)$ and put
\begin{equation}
\label{low:labelled-point}
 p(b,s)=\lambda A(b)+\varepsilon_s\mathbf{1}.
\end{equation}

At every base point $b\notin\{b_1,\ldots,b_k\}$, include all $m$ points $p(b,s)$.
At a selected terminal $b_i$, include $p(b_i,s)$ exactly when $s_i=1$.
Also include the target
\begin{equation}
\label{low:target-point}
 x_i=\lambda A(b_i).
\end{equation}
Their union is $P_{h,t,k}$.
Every coordinate is rational, with the explicit formulas
\[
 p(b,s)=\frac{(Q-1)A(b)+2(s+1)\mathbf{1}}Q,
 \qquad
 x_i=\frac{(Q-1)A(b_i)}Q.
\]
The new root $o$ is not a selected terminal, so it carries every label.
For each $s$, define the witness
\[
 w_s=p(o,s).
\]
The empty-subset witness has offset $\varepsilon_0=2/Q$.
All labelled points have positive offsets, whereas the targets have offset zero.

Let $G$ be the unit-cube graph of $P_{h,t,k}$.
Thus distinct centres $p,q$ are adjacent exactly when $\|p-q\|_\infty\le1$.
For a constructed centre, its \emph{base point} is the point $b$ used in its definition.

\begin{lemma}
\label{low:all-edges}
The points of $P_{h,t,k}$ are pairwise distinct, and their adjacencies in $G$ are as follows.
\begin{enumerate}
\item All centres at the same base point form a clique.
\item Between adjacent base points, two centres are adjacent exactly when their offsets agree.
\item There is no edge between distinct nonadjacent base points.
\item The neighbours of a target $x_i$ are exactly the retained labelled points at $b_i$.
\end{enumerate}
\end{lemma}

\begin{proof}
Regard a target as a centre with offset zero.
All offsets lie in
\[
 \left\{0,\frac2Q,\frac4Q,\ldots,\frac{2m}Q\right\},
 \qquad \frac{2m}Q<\frac14.
\]
At the same base point, two centres differ by a multiple of $\mathbf{1}$.
Their $\ell_\infty$ distance is the absolute difference of their offsets, which is less than~$1/4$.
Their offsets are different, so the centres are distinct and adjacent.

Next consider adjacent base points $b,c$.
Their difference is one of the six unit grid directions.
By Lemma~\ref{low:parity-map}, every coordinate of $A(b-c)$ is $1$ or $-1$,
and both signs occur.
If the offsets of two centres differ by $\delta$, their distance is therefore
\begin{align*}
 \|\lambda A(b-c)+\delta\mathbf{1}\|_\infty
 &=\max\{|\lambda+\delta|,|{-\lambda}+\delta|\}\\
 &=\lambda+|\delta|.
\end{align*}
For equal offsets this is $\lambda=1-1/Q<1$.
For unequal offsets we have $|\delta|\ge2/Q$, giving distance at least
$1+1/Q>1$.
This proves the second assertion.
It also shows that centres at adjacent base points are distinct.

Now let $b,c$ be distinct nonadjacent base points.
They satisfy the parity condition of Lemma~\ref{low:parity-map}, by Lemma~\ref{low:stem}.
Since they are not adjacent in the base tree, their difference is not a unit grid direction.
Thus $\|A(b-c)\|_\infty\ge2$.
The reverse triangle inequality gives
\begin{align*}
 \|\lambda A(b-c)+\delta\mathbf{1}\|_\infty
 &\ge 2\lambda-|\delta|\\
 &\ge 2\lambda-\frac{2m}Q\\
 &=2-\frac{2(m+1)}Q=\frac74>1.
\end{align*}
These centres are distinct and nonadjacent, proving the third assertion.

Finally, offset zero occurs only at the selected terminal base points $b_i$.
Every $b_i$ is twice a vertex of $H_h$, so all its coordinates are even.
Two distinct selected terminals cannot differ by a unit grid direction and hence are not adjacent
in the base tree.
A target therefore has no equal-offset centre at an adjacent base point.
By the second and third assertions, it has no neighbour at any other base point.
By the first assertion, it is adjacent to every retained labelled point at its own base point.
This proves the fourth assertion and completes the classification.
\end{proof}

\subsection{Exact distances and shattering}
\label{low:distance-section}

Let
\[
 D=6(2^h-1)+t,\qquad r=D+1,\qquad X=\{x_1,\ldots,x_k\}.
\]
We now determine the distance from every witness to every target.

\begin{lemma}
\label{low:exact-distance}
For every $0\le s<2^k$ and every $i\in[k]$,
\[
 \operatorname{dist}_G(w_s,x_i)=
 \begin{cases}
 D+1,&s_i=1,\\
 D+2,&s_i=0.
 \end{cases}
\]
\end{lemma}

\begin{proof}
The unique path in the base tree from $o$ to $b_i$ has length $D$.
No internal vertex of this path is a selected terminal, since every selected terminal is a leaf.
Thus every internal base point, as well as $o$, carries all labels.

Suppose first that $s_i=1$.
Then $p(b_i,s)$ is present.
Follow the base-tree path from $o$ to $b_i$, keeping label $s$ at every base point.
By Lemma~\ref{low:all-edges}, each consecutive pair of centres is adjacent.
This gives a path of length $D$ from $w_s$ to $p(b_i,s)$.
One further edge reaches $x_i$.
Hence $\operatorname{dist}_G(w_s,x_i)\le D+1$.

For a lower bound, consider any path in $G$ from $w_s$ to $x_i$.
Project each centre to its base point.
By Lemma~\ref{low:all-edges}, each edge either projects to an edge of the base tree
or leaves the base point unchanged.
Delete all unchanged steps from the projected walk.
The remaining walk joins $o$ to $b_i$ in the base tree, so it uses at least $D$ edges.
The last edge of the original path is inside the clique at $b_i$,
because every neighbour of $x_i$ has base point $b_i$.
Every path from $w_s$ to $x_i$ therefore has length at least $D+1$.

Suppose that such a path has length $D+1$.
It must use exactly $D$ edges between base points and exactly one edge within a base point.
The latter is the last edge.
Every preceding edge preserves the offset, by Lemma~\ref{low:all-edges}.
The path starts with the positive offset $\varepsilon_s$, so its penultimate vertex must be
$p(b_i,s)$.
This point exists exactly when $s_i=1$.
We conclude that the distance is $D+1$ when $s_i=1$, and is at least $D+2$ when $s_i=0$.

It remains to give a path of length $D+2$ in the latter case.
Choose a label $s'$ with $s'_i=1$.
Since $s_i=0$, the labels $s$ and $s'$ are different.
Both root centres are present, and one edge inside the root clique joins $w_s$ to $w_{s'}$.
Follow label $s'$ for $D$ steps to $p(b_i,s')$, then take the final edge to $x_i$.
The resulting path has length $D+2$, proving the formula.
\end{proof}

By Lemma~\ref{low:exact-distance}, the closed radius-$r$ neighbourhood of $w_s$ satisfies
\[
 N_G^r[w_s]\cap X=\{x_i:s_i=1\}.
\]
As $s$ ranges from $0$ to $2^k-1$, these intersections give every subset of $X$.
All witnesses and targets belong to the same point set $P_{h,t,k}$, and every witness uses
the same radius $r=6(2^h-1)+t+1$.
Thus the radius-$r$ neighbourhood system of $G$ has VC-dimension at least~$k$.

\subsection{The bound for every integer radius}
\label{low:all-radii}

\begin{proof}[Proof of Theorem~\ref{thm:lower}]
Let $r\ge7$ be an integer, and choose
\[
 h=\left\lfloor\log_2\frac{r+5}{6}\right\rfloor,
 \qquad t=r-(6\cdot2^h-5),\qquad k=8^h.
\]
The assumption $r\ge7$ gives $(r+5)/6\ge2$, so $h\ge1$.
The inequality $2^h\le(r+5)/6$ gives $t\ge0$.
All three parameters are integers, and they satisfy
\[
 6(2^h-1)+t+1=r.
\]
The preceding construction therefore gives a finite rational point set whose radius-$r$
neighbourhood system shatters $k=8^h$ targets.
Consequently,
\[
 f^*(r)\ge8^h
 =8^{\left\lfloor\log_2((r+5)/6)\right\rfloor}.
\]
The definition of $h$ also gives $(r+5)/6<2^{h+1}$.
Hence $2^h>(r+5)/12$, and cubing this inequality yields
\[
 f^*(r)\ge8^h>\frac{(r+5)^3}{1728}.
\]
In particular, choosing $t=0$ gives $f^*(6\cdot2^h-5)\ge8^h$ for every $h\ge1$.
\end{proof}

\subsection{Size of the construction}
\label{low:size}

For completeness, we count the points in the construction.
Let $E_h$ be the number of unit grid edges of $H_h$.
The skeleton has $2a$ edges in its $x$-segment, $4a$ edges in its two $y$-segments,
and $8a$ edges in its four $z$-segments.
The attached copies share no edges with the skeleton or with one another.
Therefore
\[
 E_0=0,\qquad E_h=14\cdot2^{h-1}+8E_{h-1}.
\]
Solving this recurrence gives
\[
 E_h=\frac73(8^h-2^h).
\]
Indeed, the formula is zero for $h=0$, and substituting it at $h-1$ gives
\begin{align*}
 14\cdot2^{h-1}+\frac{56}{3}(8^{h-1}-2^{h-1})
 &=\frac73\,8^h-\frac{14}{3}\,2^{h-1}\\
 &=\frac73(8^h-2^h).
\end{align*}
The tree $H_h$ has $E_h+1$ vertices.
Subdivision adds one vertex for each edge, and the root path adds $t$ vertices.
Thus
\[
 |B_{h,t}|=2E_h+1+t
 =1+\frac{14}{3}(8^h-2^h)+t.
\]
Every unselected base point contributes $m=2^k$ centres.
At a selected terminal, exactly $m/2$ labels have the required bit equal to~$1$,
and there is one additional target.
The total number of centres is consequently
\begin{align*}
 |P_{h,t,k}|
 &=m(|B_{h,t}|-k)+k(m/2+1)\\
 &=2^k\left(|B_{h,t}|-\frac{k}{2}\right)+k.
\end{align*}

\end{document}